\documentclass{SciPost}

\hypersetup{
	colorlinks,
	linkcolor={red!50!black},
	citecolor={blue!50!black},
	urlcolor={blue!80!black}
}

\DeclareSymbolFont{usualmathcal}{OMS}{cmsy}{m}{n}
\DeclareSymbolFontAlphabet{\mathcal}{usualmathcal}

\fancypagestyle{SPstyle}{
	\fancyhf{}
	\lhead{\colorbox{scipostblue}{\bf \color{white} ~SciPost Physics Proceedings }}
	\rhead{{\bf \color{scipostdeepblue} ~Submission }}
	
	\fancyfoot[C]{\textbf{\thepage}}
}

\usepackage{amsmath}
\usepackage{amssymb}

\newtheorem{definition}{Definition}[section]

\numberwithin{equation}{section}

\usepackage{amsmath}
\usepackage{amssymb}

\begin{document}

\pagestyle{SPstyle}

\begin{center}{\Large \textbf{\color{scipostdeepblue}{
				%%%%%%%%%% TODO: Write your article's title here
A New Definition of Quantum Superposition
%				Article Title, as descriptive as possible, ideally fitting in two lines (approximately 150 characters) or less\\
}}}\end{center}

\begin{center}\textbf{
		% Use (full) first name (+ middle name initials) + surname format.
		% Separate subsequent authors by a comma, omit comma and use "and" for the last author.
		% Mark the corresponding author(s) with a superscript symbol in this order
		% \star, \dagger, \ddagger, \circ, \S, \P, \parallel, ...
		Stephen Bruce Sontz
%		Aah B. Cee\textsuperscript{1$\star$},
%		Dee E. Faa\textsuperscript{2} and
%		Gee K. See\textsuperscript{3$\dagger$}
}\end{center}

\begin{center}
	%{\bf 1} 
	Centro de Investigaci\'on en Matem\'aticas, A.C.
	\\
	Guanajuato, M\'exico
%	\\
%	{\bf 2} Affiliation 2
%	\\
%	{\bf 3} Affiliation 3
	%%%%%%%%%% END TODO: AFFILIATIONS
	% Provide email address of corresponding author(s)
	\\[\baselineskip]
	%$\star$ 
	\href{mailto:email1}{\small sontz@cimat.mx}
	%\,,\quad
	%$\dagger$ \href{mailto:email2}{\small email2}
\end{center}

\definecolor{palegray}{gray}{0.95}
\begin{center}
	\colorbox{palegray}{
		\begin{tabular}{ll}
			\hspace{-11pt} \begin{minipage}{0.18\textwidth}
				\includegraphics[width=28mm]{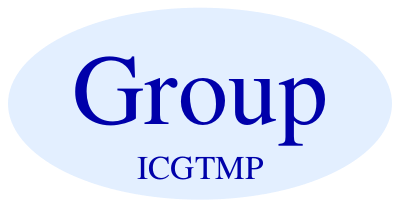}
			\end{minipage}
			&
			\begin{minipage}{0.75\textwidth}
				\begin{center}
					{\it 36th International Colloquium on Group Theoretical Methods in Physics}\\
					{\it Valladolid, 13-17 July 2026} \\
					\doi{10.21468/SciPostPhysProc.?}\\
				\end{center}
			\end{minipage}
		\end{tabular}
	}
\end{center}

\section*{\color{scipostdeepblue}{Abstract}}
\textbf{\boldmath{%
		%%%%%%%%%% TODO: frame
%		The abstract is in boldface, and should fit in 8 lines. It should be written in a clear and accessible style, emphasizing the context, the problem(s) studied, the methods used, the results obtained, the conclusions reached, and the outlook. You can add a table contents, recommended if your paper is more than 6 pages long.
The usual description of the superposition of two 
(pure quantum) states is ambiguous, since the
binary operation of addition in a Hilbert space
does not pass down to 
the quotient projective space. 
In the setting of 
complex projective space  
this paper gives a new geometrical definition 
of the 
superposition of 
two pure states, viewed as two distinct points
in the projective space, as the unique (complex) line
on which those two points lie. 	
Using the Erlangen program the geometry appropriate for quantum theory 
of the projective space
is precisely defined.  	
Also this rigorous approach is compared with others.  
}}

\vskip 0.2cm \noindent
{\bf Keywords: quantum superposition, 
geometric quantum theory, quantum state expansion}

\vskip 0.2cm \noindent
{\bf MSC2020 codes: 81P05, 81P16}

\vspace{\baselineskip}

%%%%%%%%%% BLOCK: Copyright information
% This block will be filled during the proof stage, and finilized just before publication.
% It exists here only as a placeholder, and should not be modified by authors.
\noindent\textcolor{white!90!black}{%
	\fbox{\parbox{0.975\linewidth}{%
			\textcolor{white!40!black}{\begin{tabular}{lr}%
					\begin{minipage}{0.6\textwidth}%
						{\small Copyright attribution to authors. \newline
							This work is a submission to SciPost Phys. Proc. \newline
							License information to appear upon publication. \newline
							Publication information to appear upon publication.}
					\end{minipage} & \begin{minipage}{0.4\textwidth}
						{\small Received Date \newline Accepted Date \newline Published Date}%
					\end{minipage}
			\end{tabular}}
	}}
}
%%%%%%%%%% BLOCK: Copyright information

\tableofcontents
	
\section{Introduction}

This paper is unusual since it centers on a
new definition without presenting any new theorems. 
However, the principle of superposition has been 
viewed as having 
a fundamental role since the earliest days of
quantum theory. 
For example, it was the topic of an entire 
chapter in Dirac's text \cite{dirac} of 1930 on 
quantum mechanics and 
is central to quantum foundations.  
Unfortunately, since then it has been used 
ambiguously.
This situation will be analyzed
and corrected here. 
While the reader may have seen examples of this 
ambiguity, I leave it to the reader to
identify and when possible correct 
these misuses rather than
single out specific 
examples in the literature for public rebuke. 
So the goal of this paper is to give a new unambiguous 
definition of superposition with the hope that
this will eliminate annoying confusions. 
And it turns out that this definition is 
based on a
simple geometric property. 

This paper is 
a continuation of the author's recent 
work \cite{sbs, sbs2} 
on the foundations of quantum theory as a
geometric theory based on a complex projective 
space, which can have either finite or 
infinite dimension. 
This projective space is a K\"ahler manifold,
even in the infinite dimensional case. 
This means that this space has these compatible
structures:  
a Riemannian metric (called the Fubini-Study metric),
a non-degenerate symplectic form and an almost
complex structure (which actually comes from a complex
structure). 
With just the symplectic structure  this projective
space becomes a full-fledged phase space which carries
a Hamiltonian flow embodying the time dependent
Schr\"odinger equation. 
The quantum probabilities and uncertainty relations 
are expressed solely in terms of the Riemannian metric. 
It seems to be important, though not generally recognized,
that there is 
this dichotomy between a Riemannian metric used for some
purely
quantum properties and a symplectic form used for a 
Hamiltononian dynamics that is closely allied to the
methodolgy of classical mechanics. 
However, superposition will be defined in terms of  
abstract synthetic projective geometry.  
For this geometric approach
see the papers 
\cite{anandan, ashtekar-schilling, 
ciagli, cirelli1, cirelli2,
gibbons, heslot, hughston, kibble, 
schilling} (with \cite{kibble} by Kibble 
in 1979 being the earliest) 
and the recent lecture notes
\cite{freed}.

The notation used here is that of references
\cite{sbs,sbs2}.
For a vector space $W$ we define
$W^{\rm x} := W \setminus \{ 0 \}$, where 
$0 \in W$ is the zero vector. 
Also, 
$\mathcal{H}$ always denotes a complex 
Hilbert space, possibly infinite dimensional,
whose 
inner product, denoted as 
$\langle \cdot , \cdot \rangle$, is anti-linear 
in the first entry and linear in the second.
To avoid trivial cases we always assume that 
$\dim_{\mathbb{C}} \mathcal{H} \ge 2$.

Section~2 motivates and introduces the new definition
of superposition. 
Section~3 discusses the relation of this definition to 
the expansion of a unit vector in an orthonormal basis. 
Section~4 defines on complex projective space the exact geometry relevant to quantum theory
according to the Erlangen program, while Section~5
gives a more general definition of superposition 
purely in terms of lattice theory together with a
comparison with an alternative approach of \cite{beltrametti}.  
Lastly, Section~6 has concluding remarks and
comments on possible future avenues for research. 

\section{The new definition}
 
To motivate the need for a new definition
note that the standard definition of 
{\em the} superposition of two distinct states is the
normalization of {\em a} linear combination of them. 
But since there are infinitely many 
linear combinations, superposition is not 
a binary operation with a uniquely defined 
state as its value 
for every such pair of states. 
This was already stated by Dirac in 1930 in \cite{dirac},
where superposition is presented as 
a foundational principle of quantum theory. 
Nonetheless, in the subsequent literature 
just one of these possible linear combinations is
often arbitrarily singled out and dubbed as being 
{\em the} superposition. 
This is an ambiguous methodology. 

First, let's recall that a {\em (pure quantum) state} is
an element in the {\em complex projective space }
$ \mathbb{CP}(\mathcal{H}):= 
\mathcal{H}^{\rm x}/\mathbb{C}^{\rm x}$,
where $\mathcal{H}$ is a Hilbert space 
(appropriately chosen to model a quantum system) 
and the multiplicative group $ \mathbb{C}^{\rm x} $ acts 
by scalar multiplication on $ \mathcal{H}^{\rm x}$. 
Equivalently, a state is a {\em ray} 
(that is, a one-dimensional
subspace) $\mathbb{C}\psi$ for some 
$0 \ne \psi \in \mathcal{H}$. 
The non-zero vector $\psi$, which is not
unique, is said to 
{\em represent the state}. 
Also define
$\pi : \mathcal{H}^{\rm x} \to \mathbb{CP}(\mathcal{H})$
to be the quotient map. 

Here is how to reformulate rigorously 
the standard approach to superposition. 
First one starts with two distinct
states, represented by 
vectors 
$\psi_{1}, \psi_{2} \in \mathcal{H}$,
both non-zero. 
These can be taken to be unit vectors, 
as is typically done in physics, but
that does not change the rest of this argument. 
Being distinct states is equivalent to 
$\mathbb{C} \psi_{1} \ne \mathbb{C} \psi_{2}$ and also
to saying that $ \psi_{1}, \psi_{2} $ are
linearly independent. 
It is not assumed here that $\psi_{1}$ and $\psi_{2}$
are orthogonal, though that extra,
and unnecessary, hypothesis is often made. 
For any pair of complex numbers 
$(a,b) \ne (0,0)$ we then have that the linear
combination 
$$
\psi_{ab} := a \psi_{1} + b \psi_{2} \ne 0
$$
represents a state. 
This implies that 
$\psi_{ab} \in 
\mathbb{C} \psi_{1} + \mathbb{C} \psi_{2} =: V$, 
a subspace of $\mathcal{H}$ of (complex) dimension $2$. 
By the linear independence of $ \psi_{1}, \psi_{2} $  
any non-zero vector in $V$ is equal to
$\psi_{ab}$ for a unique pair of complex numbers 
$(a,b) \ne (0,0)$. 
However, the state represented by a non-zero 
vector in $V$ does not uniquely determine the pair
$(a,b)$. 
Rather the states are the orbits of 
$\mathbb{C}^{\rm x}$ acting on 
$V^{\rm x}$
by scalar multiplication on the subset $V^{\rm x}$ 
of $V$. 
Of course, $V^{\rm x}/ \mathbb{C}^{{\rm x}}$ is 
a one-dimensional complex projective space, which 
is also called a {\em complex projective line}. 
This space, considered as a real $C^{\infty}$ 
manifold, is diffeomorphic to the standard unit sphere
in $\mathbb{R}^{3}$. 
Since this sphere is 2-dimensional as a real 
manifold, it may be difficult to think of it as
a 1-dimensional complex manifold. 
In certain contexts in quantum physics this is known as the
{\em Bloch sphere}.  
Also, this complex line is a compact topological 
space, and unlike a one-dimensional 
real line, which is isomorphic to $\mathbb{R}$ as
an affine space, 
there is no natural way to introduce on it 
arithmetic operations nor
a natural total order. 
On the other hand $\mathbb{R}$ is not compact. 

With all of the previous 
as motivation a preliminary definition of the space of
all superpositions of two linearly independent vectors 
$\psi_{1}, \psi_{2} \in \mathcal{H}$ is the
complex projective subspace 
$(\mathbb{C} \psi_{1} + \mathbb{C} \psi_{2})^{\rm x} / 
\mathbb{C}^{\rm x}$ of $\mathbb{CP}(\mathcal{H})$. 
This definition relies on the structure of the
Hilbert space used to define the projective space and so
that is why it is flagged as preliminary. 

Now denote by $ x_{1}$ (resp. $x_{2}$) the
point in the complex projective space that $\psi_{1}$ 
(resp. $\psi_{2}$) projects down to by applying $ \pi $.
Since $\psi_{1}$ and $\psi_{2}$ are linearly independent,
it follows that $x_{1} \ne x_{2}$.  
Then note that 
$$
x_{1}, x_{2} \in 
(\mathbb{C} \psi_{1} + \mathbb{C} \psi_{2})^{\rm x} 
/\mathbb{C}^{\rm x}
= V^{\rm x} /\mathbb{C}^{\rm x}
$$
and therefore $ V^{\rm x} /\mathbb{C}^{\rm x}$ 
is the unique (complex) 
line on which both $x_{1}$ and $x_{2}$ lie. 
Recall that one of the basic axioms of abstract,
synthetic projective 
geometry 
(see \cite{holland}) is the following:

\vskip 0.2cm \noindent
{\bf Axiom:} 
Any pair of two distinct points lie on 
a unique line. 

\vskip 0.2cm \noindent
This property is indeed satisfied in the specific case 
of complex projective spaces as well as for other
geometries. 
Given this motivating discussion, 
here is the rigorous definition of
superposition 
for {\em any} abstract, synthetic projective space. 

\begin{definition}
The {\em superposition} of two distinct points $x,y$
in a
projective space is the unique line on which both
points lie. 
We denote that line as $x \circledcirc y$. 
\end{definition}

In quantum theory this definition translates to saying 
that the 
{\em quantum superposition}
 of two distinct pure states is a 
{\em set} of mathematical objects, namely 
all pure states on the unique (complex) line on which
those two points lie or, equivalently, 
as one unique
mathematical {\em object}, namely that line itself. 
However, the notation favors the second option.
This latter viewpoint leads to the diagram
$
    ( \mathcal{G}_{0} \times \mathcal{G}_{0} )
    \setminus \Delta 
    \stackrel{\circledcirc}{\longrightarrow} 
    \mathcal{G}_{1}, 
$
where  $\mathcal{G}_{k}$ denotes the projective
Grassmannian of all $k$-dimensional complex
projective subspaces for every integer
$ k \ge 0$. 
Moreover, $\Delta$ denotes the diagonal subset of 
all pairs $(x,x)$ with $x \in \mathcal{G}_{0}$. 
In particular, 
$ \mathcal{G}_{0} = \mathbb{CP}(\mathcal{H})$.

This new definition does present superposition as 
some 
sort of binary operation, but with its values 
being lines. 
Explicitly, to each pair of two {\em distinct} 
points ($\equiv$ states) superposition assigns a line. 
However, a binary operation in the common usage
means an assignment for {\em any} pair. 

The next definition is as old as geometry itself, 
but a new terminology is given to emphasize its meaning
in quantum theory. 
\begin{definition}
We say that three distinct points in a projective
space are {\em colinear} or {\em superposed} if they all lie on
one (hence unique) line. 
We also say the any one of these points is 
{\em a superposition} of the other two points. 
\end{definition}

\section{Expansion of a State}

Superposition combines two states
into something else, that is not a uniquely 
defined state. 
This is often conflated with the expansion 
of one state as a combination 
of other states, possibly just  
two other states. 
This is typically only done in the setting of
a Hilbert space $\mathcal{H}$ with a given 
orthonormal basis, say 
$\{ e_{\alpha} \,|\, \alpha \in A \}$ with 
$ A $ being an index set. 
As before, we avoid trivialities by requiring that
$\dim_{\mathbb{C}} \mathcal{H} \ge 2$ or equivalently 
by requiring $A$ to have cardinality at least $2$. 
Then for any unit vector $\psi \in \mathcal{H}$, 
which represents a state,  
there are unique complex numbers 
$\lambda_{\alpha}$ such that 
\begin{equation}
	\label{expand-psi}
\psi = \sum_{\alpha} \lambda_{\alpha} e_{\alpha}
\qquad {\rm and} \qquad 
\sum_{\alpha} |\lambda_{\alpha}|^{2} = 1. 
\end{equation}
In fact, we have 
$ \lambda_{\alpha} = \langle e_{\alpha}, \psi \rangle$. 
This expansion of $\psi$ can be incorrectly  
turned around to saying that 
$\psi$ is the 
superposition of the $e_{\alpha}$'s. 
We shall now analyze this last assertion more carefully 
in order to see what goes wrong. 

First, the idea is to view the expansion 
\eqref{expand-psi} 
of $\psi$
in the complex projective space 
$\mathbb{CP}(\mathcal{H})$.
Put $x:= \pi (\psi)$ and 
$y_{\alpha}:= \pi (e_{\alpha})$. 
Then $\lambda_{\alpha}$ is not a function of $x$ and 
$y_{\alpha}$, but 
$ |\lambda_{\alpha}|^{2} = 
|\langle e_{\alpha}, \psi \rangle|^{2}$ is 
a function of $x$ and $y_{\alpha}$ provided they 
satisfy
the above conditions. 
So we define $p(y_{\alpha}, x) := 
|\langle e_{\alpha}, \psi \rangle|^{2}$
with the consequence that
$$
0 \le p(y_{\alpha}, x) \le 1 \qquad {\rm and}
\qquad   \sum_{\alpha} p(y_{\alpha}, x) =1. 
$$
It is not helpful to say that the one state $x$ 
{\em is} a superposition of the many\footnote{In this context `many' means: at least $2$.}
states $y_{\alpha}$
with each $y_{\alpha}$ having probability
$p(y_{\alpha},x)$. 
This is so, since $x$ can not in general 
be uniquely recovered from this data. 
In fact, the only cases where $x$ can be recovered are
when $ p(y_{\beta}, x) = 1$ for some $\beta \in A$
and consequently $p(y_{\alpha}, x) = 0$ for
all $\alpha \ne \beta$. 
Such cases could be called {\em classical} with respect to
the family $\{ y_\alpha  \,|\, \alpha \in A \}$. 

The idea here is that one wants to replace the state $x$
with a 
classical statistical mixture of the states 
$y_{\alpha}$ with the justification that something has 
happened to the system (say, a measurement was made) 
so that
the one state $x$ {\em transforms to} (rather than
{\em is}) 
this mixture of states. 
However, such a mixture is quite different from a 
superposition as it has been defined here or even
as it is usually understood. 
In fact for a given $x$ this mixture allows one to 
define a probability measure 
on $\mathbb{CP}(\mathcal{H})$ whose 
support is a subset of  
$\{ y_{\alpha} \,|\, \alpha \in A \}$. 
And any probability measure 
on $\mathbb{CP}(\mathcal{H})$, the quantum phase space, 
can be understood in the same manner as any 
probability measure on a classical mechanical phase
space, that is, as a 
classical statistical mixture of states or, 
more simply put,
as a {\em mixed state}. 
And according to spectral theory 
(see \cite{weid})
such mixtures 
as described here are in 
bijective correspondence with the usual definition
of a mixed state\footnote{Also called a density matrix.}
 as a positive trace class operator
with trace $1$. 

As an aside, note that these mixtures are just a special 
case of an arbitrary probability measure on 
$\mathbb{CP}(\mathcal{H})$, which might be the 
appropriate structures to be considered in quantum
statistical mechanics, rather than the limited space 
of density matrices. 

The upshot is that given an 
orthonormal basis 
$\{ e_{\alpha} \,|\, \alpha \in A \}$
the pure state represented by 
$\psi$ is replaced by the mixed state 
$$
\sum_{\alpha} 
|\langle e_{\alpha}, \psi \rangle|^{2} \,
|e_{\alpha} \rangle \langle e_{\alpha}| = 
\sum_{\alpha} 
p(y_{\alpha}, x)  \,
|e_{\alpha} \rangle \langle e_{\alpha}|,
$$
where the Dirac ket-bra notation has been used. 
While, as already noted, this can be given a 
meaning in the projective space, it is not enough 
in general to recover $x$. 

Of course, one is perfectly justified in
using expansions such as 
\eqref{expand-psi} as a mathematical tool in the 
analysis of a quantum system. 
But one should be careful never to use language that 
ascribes to this formula any extra meaning.  

A common example is a $2$-dimensional Hilbert space
with orthonormal basis $\{ \psi_{u}, \psi_{d} \}$. 
One might think of this as a simple model of a
particle with spin $1/2$, where with respect to a 
vertical axis $\psi_{u}$ means a particle whose
spin measured along that axis always yields
the spin pointing up,
whereas $\psi_{d}$ means the same measurement 
always yields the spin pointing down. 
But suppose a sequence of measurements of 
such particles, which are identically prepared 
to have the same spin along another axis, yield 
spin up with relative frequency $1/2$. 
Consequently, this yields spin down with relative
frequency $1/2$, since these are exclusive and
exhaustive outcomes. 
So a standard, but erroneous, 
way to describe this is to say that
the particles were necessarily 
all prepared in the superposed state 
represented by 
$$
   \psi_{sp}:= \dfrac{1}{\sqrt{2}} \psi_{u} +
   \dfrac{1}{\sqrt{2}} \psi_{d},
$$
since this state has probabilities
$|\langle \psi_{sp} , \psi_{u} \rangle|^{2} =1/2$ and 
$|\langle \psi_{sp} , \psi_{d} \rangle|^{2} =1/2$, 
which correspond to the observed frequencies. 
However, the superposed states 
represented by 
$$
\psi_{sp, \theta}:= \dfrac{1}{\sqrt{2}} \psi_{u} + 
e^{ {\rm  i} \theta }
\dfrac{1}{\sqrt{2}} \psi_{d} \qquad 
{\rm where~} \theta \in [0, 2 \pi)
$$
form an uncountably infinite family of unit vectors 
representing pairwise 
distinct states with the same probabilities, 
namely 
$|\langle \psi_{sp,\theta} , \psi_{u} \rangle|^{2} =1/2$ and 
$|\langle \psi_{sp,\theta} , \psi_{d} \rangle|^{2} =1/2$. 
Actually, the system could have been prepared 
in any combination
of the states $ \psi_{sp,\theta} $ for various different
values of $ \theta $ with identical resulting frequencies 
of the vertical spins, namely $1/2 $ for each result. 
As noted earlier, this exemplifies that 
probabilities alone do not determine a unique 
corresponding superposed state or even the
existence of such a state. 

This misunderstanding of superposition can occur
with any Hilbert space $\mathcal{H}$ with
$\dim_{\mathbb{C}} (\mathcal{H}) \ge 2$, where now
$\{ \psi_{u}, \psi_{d} \}$ is any orthonormal subset
of $ \mathcal{H}$. 
For example, consider the case 
$\dim_{\mathbb{C}} (\mathcal{H}) = \infty$. 
Then $\psi_{sp, \theta}$ (and in particular
$\psi_{sp} = \psi_{sp, 0}$) can be expanded in any
orthonormal basis while 
$\psi_{u}$ and $\psi_{d}$ are each expanded in 
other distinct orthonormal bases. 
This may lead to a computational nightmare,
but the point is the the same probabilities would
be the results as obtained above. 
However, any one particular expansion of a unit
vector is not unique and therefore is not 
{\em the} superposition of it, albeit that it
is convenient for doing some calculations.

\section{The Geometry of $\mathbb{CP}(\mathcal{H})$}

It is a little too facile to just say that we 
are working with the geometry of complex 
projective space. 
Exactly, what geometry are we using? 
The answer depends on the choice of a symmetry group,
since geometric properties are those that are
invariant under the action of that group, 
according to the Erlangen program. 
One group commonly used is the 
projective general 
linear group, also called the homography group,
$ PGL(\mathcal{H}) := GL(\mathcal{H})/ Z(GL(\mathcal{H})) $ 
where $ GL(\mathcal{H})$ is defined to be 
the group of all bounded linear 
invertible operators $ T: \mathcal{H} \to \mathcal{H} $ and
$ Z(GL(\mathcal{H}))$ is its center. 
Another possibility is the group of collineations, 
which is strictly larger. 
These are the standard choices for defining the
geometry of $\mathbb{CP}(\mathcal{H})$.
Either of these two options will be called,
rather ambiguously, the standard geometry. 

However, in the context of quantum theory we look for
a group that preserves only probabilities. 
The simplest probability is given by Born's Rule  
$|\langle \phi, \psi \rangle |^2$ where  
$ \phi, \psi \in \mathcal{H} $ are unit vectors. 
This is clearly a function of $\pi(\phi)$
and $\pi(\psi)$. 
By Wigner's theorem any transformation of 
$\mathbb{CP}(\mathcal{H})$ that preserves 
these probabilities lifts via the quotient map $ \pi $ to a 
unitary or an anti-unitary transformation of $ \mathcal{H}$, 
since $ \dim_\mathbb{C} \mathcal{H} \ge 2$.
(See \cite{wigner}.) 
It seems that there is no widely accepted name or
notation for the group of all 
unitary and anti-unitary transformations of $ \mathcal{H}$.
I will denote this group as $U^\pm (\mathcal{H}) $. 
It contains the unitary group $U (\mathcal{H}) $
as an index 2 subgroup. 
So the projectivization $PU^\pm (\mathcal{H}) $
of $U^\pm (\mathcal{H}) $ might be considered
to be the correct group defining the geometry relevant 
to the quantum theory of
$\mathbb{CP}(\mathcal{H})$. 
With either choice the definitions of point and
line are the same and so the Axiom introduced above is
valid. 
Note that 
with either choice we obtain a geometry
for quantum theory that is distinct from
that associated with $  PGL(\mathcal{H}) $.

But the mathematical operation of time reversal, 
which is anti-unitary, 
is not universally a symmetry in physics. 
Since this paper is intended to be a foundation for all
of quantum physics, I opt
to postulate the projectivization 
$PU (\mathcal{H}):= U(\mathcal{H})/U(1) $ 
of $U (\mathcal{H}) $ as the
correct group in quantum theory for defining the geometry of 
$\mathbb{CP}(\mathcal{H})$. 
It is important 
that $ PU (\mathcal{H}) $ preserves  
the quantum consecutive
probability and quantum conditional probability as
given in \cite{sbs-qm-book} as can be easily shown. 
Further justification for this choice is given in 
\cite{freed}. 

The fact that 
$ PU (\mathcal{H}) \subset PGL(\mathcal{H})$ is a proper
inclusion implies that
every standard geometric property of $\mathbb{CP}(\mathcal{H})$
is a quantum property, while there are quantum properties 
of  $\mathbb{CP}(\mathcal{H})$
that are not standard geometric properties. 
This latter type of quantum properties merits
further investigation;  
some of these have already appeared in \cite{sbs2},
though they first appeared in \cite{bargmann} and are
therefore called {\em Bargmann invariants}. 
Also see \cite{prataps} and references therein for 
more details and applications. 

\section{Superposition as a Lattice Property}

The definition of
superposition of two distinct states
($\equiv$ zero dimensional projective subspaces), 
as the unique line 
($\equiv$ one dimensional projective subspace) 
on which they lie, can be rephrased as being the
smallest closed projective subspace containing them. 
This motivates extending the 
definition of
superposition to any cardinal number
of states, or even of closed complex projective subspaces, 
by defining it as follows:
\begin{definition}
Let $ \{ S_\alpha\}_{\alpha \in A}  $ be any non-empty
family of closed projective subspaces. 
Then its {\em superposition} is the smallest closed 
projective subspace containing every $ S_\alpha$ 
or in other words its {\em supremum} 
$ \vee_{\alpha \in A} \, S_\alpha $. 
 
\end{definition}
As such, superposition is a lattice property, 
namely the supremum, 
associated to the projective geometry. 
With this extended definition the superposition of a 
state $ x $ with itself is the same state $ x $. 
In particular,
superposition of two states becomes a binary operation, 
but now with codomain
$\mathcal{G}_0 \cup \mathcal{G}_1 $.  

Also, in accordance with our general philosophy, 
superposition has now 
been defined again in purely
geometric terms without any reference to 
Hilbert space theory. 
In fact, it clearly generalizes to any mathematical  
structure that has suprema, such as the lattice
of closed complex projective subspaces that 
corresponds bijectively with the projections of
any von Neumann algebra given as an algebra of 
operators acting on $ \mathcal{H}$. 

A different mathematically rigorous definition 
of superposition 
is given by the authors of \cite{beltrametti}. 
Actually, they work in a more
general setting than lattice theory.
They start with a partially ordered set 
$\mathcal{L}$ that
has both a maximal element, denoted $I$, and a
minimal element, denoted $0$. 
Furthermore, they require for every countable
disjoint family of elements, 
namely $x_j \in \mathcal{L} $
and $\inf \{ x_j, x_k\} = 0  $ for all $j \ne k $, that
$ \sup_j x_j $ exists. 
Then they define a probability measure on 
$\mathcal{L}$ to be a function 
$p : \mathcal{L} \to [0,1]$ that satisfies 
the normalizations 
$ p(0) = 0 $ and $p(I) = 1 $ as well as
$ \sigma $-additivity, namely 
$$
    p ( \sup x_j) = \sum_j p(x_j) 
$$
for every disjoint countable family $x_j \in \mathcal{L} $. 
This is in the spirit of 
Gleason's theorem 
(see \cite{gleason,varadarajan}) which shows
in the setting of a Hilbert space $ \mathcal{H}$ 
of dimension $ \ge 3$
that such probability
functions have a unique representation in 
terms of density matrices, which in turn are
physically understandable as mixed states. 
The set $ $$\mathcal{P}$
of all probability measure on $ \mathcal{L} $
then can be shown to be convex. 
(Actually, it is $ \sigma $-convex. See \cite{beltrametti}.)
In resonance with the theories of measure theoretic probability and of $ C^*$-algebras 
these probability measures are identified
in \cite{beltrametti} as the
states of their theory, while $ \mathcal{L} $
plays the role of a lattice of events, which
therefore corresponds to the lattice of closed
complex projective spaces used here. 

They continue with two auxiliary definitions. 
Say $ \mathcal{S} $ is any non-empty set of $ \mathcal{P} $
and $X$ is any non-empty subset of $ \mathcal{L}  $. 
Then they define 
$$
\mathcal{S}_1 (X) := \{ \mu \in \mathcal{S} ~|~ \mu(x) = 1 
\mathrm{~for~all~} x  \in X
 \}
$$
and
$$
\mathcal{L}_1 (\mathcal{S}) := \{ x \in \mathcal{L} ~|~ \mu(x) = 1
\mathrm{~for~all~} \mu \in \mathcal{S} \}. 
$$
This suffices for giving their definition 
of superposition which appears in \cite{beltrametti}. 
\begin{definition}
[Beltrametti, Cassinelli]
One says 
a probability measure $ \mu \in \mathcal{P}$ is 
{\em a superposition} of the elements
$\mathcal{S} $, a subset of $ \mathcal{P} $,  if 
$ \mathcal{L}_1(\mathcal{S}) $ is non-empty and 
$\mu \in \mathcal{S}_1 (  \mathcal{L}_1(\mathcal{S}) ) $.
\end{definition}
As noted in \cite{beltrametti} this definition reduces
to what one wants in the case when $ \mathcal{L} $ is the
lattice of all closed subspaces of $ \mathcal{H} $. 
The motivation in \cite{beltrametti} is both to find
general structures underlying standard quantum theory
as well as to look for possible generalizations of it. 
That is what is behind this definition 

However, 
in the approach of this paper 
superposition is a 
property of closed complex projective subspaces, or 
equivalently of events 
(including states) with probability
measures playing no role. 
Then states as used in this paper correspond 
to elements of $ \mathcal{L}$,
not of $ \mathcal{P} $ as 
in the approach of \cite{beltrametti}. 
Thus their approach differs from that presented here.
More details on this alternative approach to
superposition can 
be found in \cite{beltrametti}. 

Also it should be noted that the two 
structures $ \mathcal{L}$ and $ \mathcal{P} $
are in some sense dual to each other, as is 
noted in \cite{beltrametti}.  
This motivates a definition dual 
to that given in \cite{beltrametti}.
We continue with the notation given above. 

\begin{definition}
An element $ x \in \mathcal{L} $ is 
{\em a superposition} of the elements 
$X \subset \mathcal{L}$ if $ \mathcal{S}_1 (X) $ 
is non-empty and 
$ x \in \mathcal{L}_1 ( \mathcal{S}_1 (X)  ) $. 
\end{definition}
Notice that
there can be many superpositions, 
just as with the previous definition.
However, if a non-empty $X$ has a supremum, denoted 
$ \vee X$, and $ \mathcal{S}_1 (X) $ is non-empty,
then we can  show that 
$ \vee X $ is a superposition of the elements
of $X$ as follows. 
First, take any probability measure 
$p \in \mathcal{S}_1 (X) $ and any $ x \in X$.  
Then $ x \le \vee X \le I $ implies 
$ 1 = p(x) \le p(\vee X) \le p (I) = 1 $.  
Thus $ p(\vee X) = 1 $ for all 
$p \in \mathcal{S}_1 (X) $. 
This in turn implies 
$ \vee X \in \mathcal{L}_1 ( \mathcal{S}_1 (X)  ) $, 
which according to the previous definition means that 
$ \vee X $ is a superposition of the elements of $X$. 
In particular, every non-empty countable 
$X$ has a superposition. 
However, this superposition need not be unique. 

This should be compared with 
using the first definition of this section but now 
in the setting of quantum theory where 
$ \mathcal{L} $ is the lattice of
closed complex projective subspaces. 
In this case  
there is a unique superposition 
for every non-empty $X \subset \mathcal{L}$, namely 
the supremum of $X$.

\section{Conclusion}

Implicitly, this discussion is concerned with the
case of a Type~I factor with no superselection 
rules. 
In a more general setting the same geometric
definition of superposition would apply 
with the caveat that the superposed states 
might not have a physical meaning. 
One example is a superposed state of two states 
which describe two particles with unequal 
electric charges. 

It can not be left unnoticed that the above Axiom 
already appeared   
some 2300 years ago
in the book 
{\em Elements} by Euclid, although in a constructive 
language that 
appears stilted to most contemporary mathematicians. 
Euclid can be paraphrased as saying 
in his first postulate
that given two distinct points
one can draw a line segment connecting them
and then in his second postulate that any line 
segment can be further extended to be 
included in a (straight) line. 
This formulation shows how geometry developed
as a way for dealing with physical, 
spatial locations and 
their interrelations as they arose in 
practical situations, such as land 
management and architecture. 
This is consistent with the etymology of
the word `geometry' though the current usage
includes many other contexts, such as dimensions greater
than $3$, 
not considered by Euclid. 

This definition of superposition 
can be generalized to any 
geometric space for which the above 
mentioned Axiom holds. 
This includes vector and affine 
spaces\footnote{The definition of line in these
cases is well known.} over any field or division ring 
(with any dimension, even infinite dimension) 
as well as many, though
not all, Riemannian manifolds, where lines are defined 
as minimal geodesics. 
If we define the lines of Minkowski space to be the
world-lines of that geometry, then the 
Axiom does not hold. 

As a path for future research note that 
all theorems of complex projective 
geometry also apply to quantum theory. 
The theorems of Pappus and of Desargues are 
just two examples. 
The complex projective Grassmanians may have
some importance, too. 
However, the goal of such research would be to 
understand their significance in the quantum context.

\section*{Acknowledgments}

I thank all the organizers of GROUP36, held in 
Valladolid, Spain in the week 13--17 
July, 2026 for the 
opportunity to present my ideas on 
quantum theory foundations, which form the
basis for this paper. 
I am most grateful to 
Mariano Antonio del Olmo Mart\'inez
who made me aware of this 
wonderful event and encouraged my 
participation in it.

\paragraph{Funding information}
This research received no funding.

\end{document}